\documentclass[amssymb,12pt,showpacs,aps]{revtex4}
\usepackage{graphicx}
\begin{document}
\title{Spectral statistics for scaling quantum graphs}
\date{\today}
\author{Yu. Dabaghian}
\affiliation{Department of Physiology,
Keck Center for Integrative Neuroscience,\\
University of California,
San Francisco, California 94143-0444, USA
E-mail: yura@phy.ucsf.edu}

\begin{abstract}
The explicit solution to the spectral problem of quantum graphs is used 
to obtain the exact distributions of several spectral statistics, such as 
the oscillations of the quantum momentum eigenvalues around the average, 
$\delta k_{n}=k_{n}-\bar k_{n}$, and the nearest neighbor separations, 
$s_{n}=k_{n}-k_{n-1}$.
\end{abstract}

\pacs{05.45.+b,03.65.Sq}
\maketitle

\section{Introduction}

Understanding statistical properties of the quantum spectra produced 
by the classically nonintegrable systems is one of the most fundamental 
problems of quantum chaos theory. According to the Random Matrix Theory
(RMT), the statistics of the spectral fluctuations is universal, and 
reflects only the general symmetry properties of system \cite{BGS}. The 
universality of the RMT approach and its ability to produce specific 
predictions are particularly valuable since the spectra of the systems 
to which it is applied are typically out of reach for direct analytical 
studies. On the other hand, understanding how the RMT distributions 
themselves emerge from the phase space structures used by semiclassical 
spectral theories, such as the periodic orbit theory, is a matter of intense 
research. There has been a number of publications \cite{Bogomolny}
dedicated to computing spectral statistics that are accessible via periodic 
orbit expansion for the density of states. This analysis was particularly 
complete for the quantum graphs \cite{Gaspard,QGT}, which are simple 
and convenient models of quantum chaos.

As a reminder, quantum graphs consist of a quantum particle moving on a
quasi one-dimensional network with $B$ bonds and $V$ vertexes. In the limit 
$\hbar =0$, these systems produce a nonintegrable (mixing) classical 
counterpart - a classical particle moving on the same network, scattering 
randomly on its vertexes.
Despite the apparent simplicity such system, its classical behavior exhibits
many familiar features of multidimensional deterministic chaotic systems 
\cite{Gaspard}, which is clearly manifested in quantum regime. Numerical
investigations of the quantum graph spectra \cite{QGT} show that the
spectral distributions of sufficiently complicated networks are closely
approximated by the RMT Wignerian distributions. On the other hand, periodic
orbit theory for quantum graphs yields exact harmonic expansions for the
density of states, spectral staircase, quantum and classical zeta functions,
etc. Moreover, as shown recently in \cite{Fabula}, the periodic
orbit theory for the quantum graphs can be ``localized'' - it is possible to
obtain the harmonic series expansion representation for the \emph{individual}
eigenvalues of the energy or of the momentum, $k_{n}=k(n)$, as a global 
function of the index $n$. Hence, despite classical nonintegrability the spectral 
problem for quantum networks is exactly solvable within the periodic orbit theory 
approach. This fact provides an interesting opportunity to obtain analytically 
several new spectral statistics in terms of the periodic orbit theory, which is 
the main subject of this Letter.

\section{Spectral distributions for regular quantum graphs}

In the simplest case of the so called regular graphs \cite{Prima,Stanza} 
the individual momentum eigenvalues can be expanded into a periodic orbit series, 
\begin{equation}
k_{n}=\frac{\pi}{L_{0}}n-\frac{2}{L_{0}}\sum_{p}\frac{A_{p}}{\omega_{p}}
\sin \left(\frac{\omega_{p}}{2}\right) \sin \left(\omega_{p}n\right).  
\label{kn}
\end{equation}
Here $L_{0}$ is the total action length of the graph, $A_{p}$ is the weight factor 
of a periodic orbit $p$, which is a function of the scattering coefficients at the 
graph vertexes \cite{Prima,Stanza}. 
The frequency $\omega_{p}=\pi\left(m_{p}^{1}L_{1}+...+m_{p}^{B}L_{B}\right)/L_{0}$, 
is defined by the numbers of times, $m^{i}_{p}$, the orbit $p$ passes over the bond 
$i$. As shown in \cite{Stanza,Fabula}, the series in (\ref{kn}) is convergent and 
bounded by $\pi /L_{0}$, so the values $k_{n}$ are locked within a sequence of periodic 
cells. The first term in (\ref{kn}) gives the average (Weyl) behavior of the eigenvalue 
$\bar{k}_{n}$ and the subsequent periodic orbit sum describes the fluctuations 
$\delta^{(0)}_{n}=L_{0}\left(k_{n}-\bar{k}_{n}\right)/\pi$ around the average.

The transition from the exact formula (\ref{kn}) to statistical description of
$\delta^{(0)}_{n}$s can be made based on the well known fact from the analytical 
number theory \cite{Karatsuba} that the sequence of the remainders $x_{n}=\left[
\alpha n\right]_{\mathop{\rm mod}1}$, $n=1,2,...$, for any irrational number -
$\alpha $ is uniformly distributed over the interval $x\in \left[0,1\right]$. 
Assuming the generic case in which every  $\Omega_{i}=L_{i}/L_{0}$ is an irrational 
number, it is clear that parsing through the spectral sequence $k_{1}$, $k_{2}$, ..., 
$k_{n}$,..., will generate a sequence of random phases in (\ref{kn}), defined via the
combinations
\begin{eqnarray}
x_{i}=\left[\pi\Omega_{i}n\right]_{\mathop{\rm mod}2\pi},
\label{xrand}
\end{eqnarray}
that are uniformly distributed in the interval $[0,2\pi]$. Hence, the deviations of the 
eigenvalues from the average, $\delta^{(0)}_{n}$, are statistically described by a 
series of random inputs
\begin{equation}
\delta^{(0)}_{x}=-\frac{2}{\pi}\sum_{p}\frac{A_{p}}{\omega_{p}}\sin
\left(\frac{\omega_{p}}{2}\right) \sin \left(\vec{m}_{p}\vec{x}\right),
\label{deltax}
\end{equation}
where $\vec{x}=\left(x_{1},...,x_{B}\right)$, which has the same structure as the 
periodic orbit sum (\ref{kn}) (see Fig.~1).
%%%%%%%%%%%%%%%%%%%%%%%%%%%%%%%%%%%%%%%%%%%%%%%%%%%%%%%%%%%%%%%%%%%%%%%%%%%%
\begin{figure}[tbp]
\begin{center}
\includegraphics{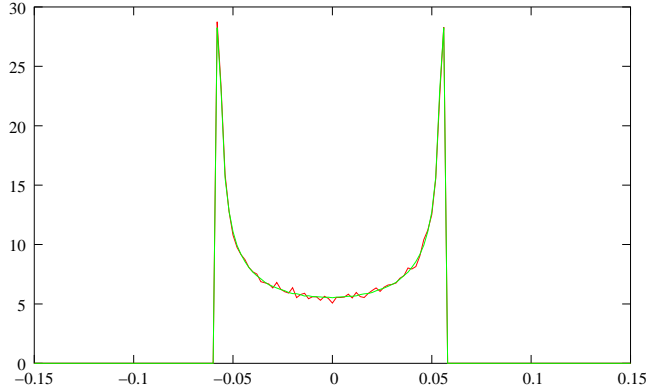}
\end{center}
\caption{The histogram of the spectral fluctuations for the simplest regular two 
bond star graph obtained from the exact expansion (\ref{kn}) (shown in red), compared 
to the distribution obtained from (\ref{deltax}) using the orbits that include up to 6 
scattering events (shown in blue).}  %%(1/pi)*asin(r)=0.0548849
\label{Fig.1}
\end{figure}
%%%%%%%%%%%%%%%%%%%%%%%%%%%%%%%%%%%%%%%%%%%%%%%%%%%%%%%%%%%%%%%%%%%%%%%%%%%
The distribution of $\delta $'s can then be obtained as 
\begin{eqnarray}
P^{(0)}_{\delta}(\delta^{(0)})=\int_{0}^{2\pi}\delta\left(\delta^{(0)}+\sum_{p}C^{(0)}_{p}
\sin\left(\vec{m}_{p}\vec{x}\right) \right)dx
\label{pdelta}
\end{eqnarray}
where $C_{p}^{(0)}=\frac{2A_{p}}{\pi\omega_{p}}\sin \frac{\omega_{p}}{2}$ and 
$dx=\prod_{i}\frac{dx_{i}}{2\pi}$. Using the exponential representation of the 
$\delta$ - functional, one finds that 
\begin{equation}
P^{(0)}_{\delta}\left(\delta^{(0)}\right) =
\int dk\,e^{ik\delta^{(0)}}F^{(0)}_{\delta}(k),
\label{distreg}
\end{equation}
where 
\begin{eqnarray}
F^{(0)}_{\delta}(k) =\int_{0}^{2\pi} e^{ik\sum_{p}C_{p}^{(0)}
\sin \left(\vec{m}_{p}\vec{x}\right)}\prod_{i}dx
\label{Fk}
\end{eqnarray}
is the characteristic function of the distribution.
As in the case of the series expansion for $k_{n}$, a finite order ($l$th)
approximation to the exact result in (\ref{distreg}) is obtained by
considering only the orbits that involve a particular number ($l$) of vertex
scatterings \cite{Prima,Stanza}.

Similarly, for the spacings between two eigenvalues, $s^{(0)}_{n,m}=k_{n+m}-k_{n}$, 
the periodic orbit expansion,
\begin{equation}
s_{n,m}^{(0)}=\frac{\pi}{L_{0}}m-\sum_{p}D_{p,m}^{(0)}\cos\left(\omega_{p}
\left( n-\frac{m}{2}\right) \right),
\label{regnn}
\end{equation}
where $D_{p,m}^{(0)}=\frac{4}{L_{0}}\frac{A_{p}^{(0)}}{\omega_{p}}\sin\left( 
\frac{\omega_{p}}{2}\right) \sin \left(\frac{\omega_{p}m}{2}\right)$, yields 
the distribution $P_{s_{m}}^{(0)}$ of the type (\ref{distreg}), where now
\begin{equation}
F^{(0)}_{s_{m}}(k)=\int_{0}^{2\pi}e^{ik\sum_{p}D_{p,m}^{(0)}\sin 
\left(\vec{m}_{p}\vec x-\frac{\omega_{p}m}{2}\right)}dx
\label{regnndist}
\end{equation}
%%%%%%%%%%%%%%%%%%%%%%%%%%%%%%%%%%%%%%%%%%%%%%%%%%%%%%%%%%%%%%%%%%%%%%%%%%%%
\begin{figure}[tbp]
\begin{center}
\includegraphics{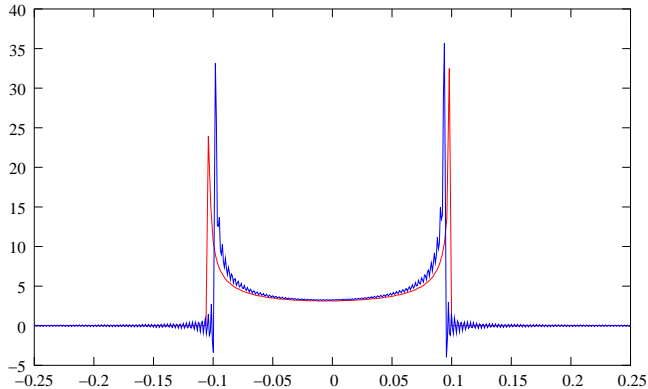}
\end{center}
\caption{The histogram of the nearest neighbor separations for the regular 
two bond star graph obtained via direct numerical solution of the spectral 
equation (shown in red), compared to the distribution obtained from
(\ref{regnndist}) (shown in blue). To emphasize the connection with Fig.~1,
the graph is scaled and shifted along the abscissa by $\pi/L_{0}$.} 
\label{Fig.2}
\end{figure}
%%%%%%%%%%%%%%%%%%%%%%%%%%%%%%%%%%%%%%%%%%%%%%%%%%%%%%%%%%%%%%%%%%%%%%%%%%%
Note, that despite the complete nonintegrability of the underlying classical 
system \cite{Gaspard}, the statistics of the spectral fluctuations of certain 
(e.g. some regular) quantum graphs is not described by the RMT and so the shapes 
of their probability distributions deviate noticeably from the familiar Wignerian 
profiles (Fig.~2).

All the above distributions are exact and depend explicitly on the specific graph 
parameters - the vertex scattering coefficients and the periodic orbit set. 

\section{Spectral hierarchy method}

As shown in \cite{Fabula}, a generic quantum graph systems is
irregular. The idea of producing the statistical distribution profiles for
irregular quantum graphs is based on three key properties of the spectral 
determinant $\Delta (k)$ \cite{Fabula}, briefly outlined below.
First, its roots as well as the roots of all of its derivatives are real.
Secondly, there is exactly one root of $\Delta^{(j)}(k)$ between every two 
neighboring roots of $\Delta^{(j+1)}(k)$ \cite{Levin}. This implies that the 
zeroes of $\Delta (k)$ can be bootstrapped by the zeroes of $\Delta'(k)$, 
while the latter sequence can be bootstrapped by the roots of $\Delta''(k)$ 
and so on (Fig.~3). Hence, there exists a hierarchy of almost periodic sequences, 
$\hat{k}_{n}^{(j)}$, $j=0,1,...$, $\Delta^{(j)}\left(\hat{k}_{n}^{(j)}\right)=0$, 
such that $\hat{k}_{n-1}^{(j)}<\hat{k}_{n}^{(j-1)}<\hat{k}_{n}^{(j)}$ and 
\begin{equation}
\hat{k}_{n}^{(j)}=\frac{\pi}{L_{0}}\left(n+\delta_{n}^{^{(j)}}\right).
\label{jsep}
\end{equation}
Lastly, the higher is the order $j$ of the derivative, the smaller are the 
fluctuations of $\delta_{n}^{(j)}$. In fact, the roots of a sufficiently 
high order $r$ of the derivative, $\Delta^{(r)}(k)$, can be locked between a 
periodic bounding sequence of points, just as the spectra of the regular graphs 
\cite{Fabula}. Hence after $r$ steps there is no need to consider the 
roots of $\Delta^{(r+1)}(k)$ to separate $k_{n}^{(r)}$ from one another. Instead, 
one can use the periodic sequence (\ref{jsep}) with 
$\delta^{(r+1)}_{n}=\frac{\pi}{L_{0}}\left(n+\frac{1}{2}\right)$, which 
explicitly carries the index $n$. The regular graphs discussed in the previous section 
correspond to the case $r=0$. 
%%%%%%%%%%%%%%%%%%%%%%%%%%%%%%%%%%%%%%%%%%%%%%%%%%%%%%%%%%%%%%%%%%%%%%%%%%%%%%%
\begin{figure}[tbp]
\begin{center}
\includegraphics{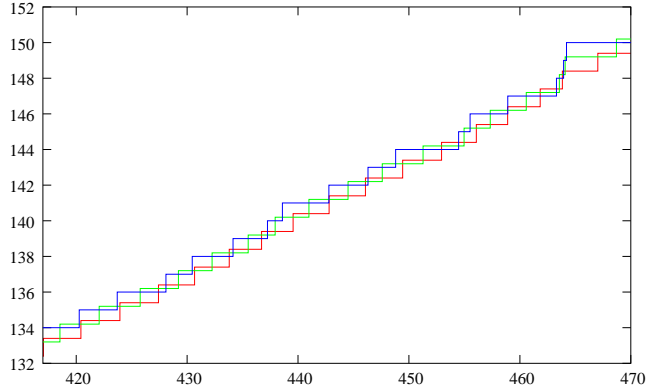}
\end{center}
\caption{The bootstrapping of the spectral staircase (shown in red) by 
the $N^{(1)}(k)$ staircase (shown in green), bootstrapped in turn by the 
$N^{(2)}(k)$ staircase (shown in blue). Small vertical shift of $N^{(1)}(k)$ 
and $N^{(2)}(k)$ is given for illustration purposes. Note that the $N^{(2)}(k)$ 
does not bootstrap the $N^{(0)}(k)$. Data obtained for the ``fully connected 
quadrangle'' graph \cite{QGT}.}
\label{Fig.3}
\end{figure}
%%%%%%%%%%%%%%%%%%%%%%%%%%%%%%%%%%%%%%%%%%%%%%%%%%%%%%%%%%%%%%%%%%%%%%%%%%%%%%%%%
These three properties of the auxiliary sequences $k^{(j)}_{n}$ allow to
obtain the whole hierarchy explicitly. Starting from the periodic separators, 
one can find the roots $\hat{k}_{n}^{(r)}$ of the $r$th derivative of spectral 
determinant, then use them as separators on the next level of the hierarchy to 
find $\hat{k}_{n}^{(r-1)}$, and so on. After $r+1$ steps,
\begin{equation}
\hat{k}_{n}^{(j-1)}=\int_{\hat{k}_{n-1}^{(j)}}^{\hat{k}_{n}^{(j)}}
\rho^{(j-1)}(k)\,kdk,
\label{seps}
\end{equation}
$j=r+1,...,1$, the spectrum $k_{n}$ is produced \cite{Fabula}.
The integrations (\ref{seps}) are made explicit by using the series expansions 
for the density $\rho^{(j)}(k)$ of the separators on each level of the hierarchy, 
\begin{equation}
\rho^{(j)}(k)=\frac{L_{0}}{\pi}+\mathop{\rm Re}
\sum_{p}S^{(j)}_{p}A_{p}^{(j)}e^{iS^{(j)}_{p}k}.
\label{densities}
\end{equation}
The weight coefficients $A_{p}^{(j)}$ and expansion frequencies $S^{(j)}$ in 
the expansions (\ref{densities}) are different for the different levels $j$ of 
the hierarchy %\cite{Tantra}. 
For $j=0$ the series (\ref{densities}) is just the usual Gutzwiller's formula.

Using (\ref{jsep}) and (\ref{densities}) in (\ref{seps}) one can define each 
element in the $j$th separating sequence in terms of the elements of the previous 
one. The result is a set of equations that completely describe the propagation of 
the fluctuations across the hierarchy,
\begin{eqnarray}
\delta_{n}^{(j-1)}=f^{(j-1)}_{\delta} -\sum_{p}C_{p}^{(j-1)}
\sin\left(\omega_{p}^{(j-1)}n+\varphi^{(j-1)}_{p}\right),
\label{deltaj}
\end{eqnarray}
where now the ``zero frequency'' term, $f^{(j-1)}_{\delta}=\left(\delta_{n}^{(j)}-
\delta_{n-1}^{(j)}-(\delta_{n}^{(j)})^{2}+(\delta_{n-1}^{(j)})^{2}\right)/2$,
the expansion coefficients, $C_{p}^{(j-1)}=\frac{2}{\pi}\frac{A_{p}^{(j-1)}}{\omega_{p}
^{(j-1)}}\sin\frac{\omega_{p}^{(j-1)}}{2}\left(\delta_{n}^{(j)}-\delta_{n-1}^{(j)}+1
\right)$, and the phases, 
\begin{equation}
\varphi_{p}^{(j-1)}=\omega_{p}^{(j-1)}\left(\delta_{n}^{(j)}+\delta_{n-1}^{(j)}-1\right)/2
\label{phi}
\end{equation}
are functions of the fluctuations $\delta_{n}^{(j)}$ and $\delta_{n-1}^{(j)}$ on the 
previous level of the hierarchy. In the particular case when $r=0$, $\delta_{n}^{(1)}=1/2$, 
$n=1$, ..., (\ref{deltaj}) coincides with the oscillating part of (\ref{kn}). 

The properties of the fluctuation sequences outlined in this Section are inherited 
by other spectral characteristics, such as the nearest neighbor separations (Fig.~4), 
etc. The corresponding systems of equations that describe the propagation of these 
characteristics across the hierarchy are similar to (\ref{deltaj}) and can be easily 
derived from relating a particular spectral sequence to the one that precedes it in 
the hierarchy via (\ref{seps}).

\section{Fluctuation Statistics}

The expansions (\ref{deltaj}) allow to produce the probability distributions for the 
spectral fluctuations across the hierarchy, using the same approach as in Section II. 
By construction, the index $n$ is the same across the hierarchy, so the combinations 
(\ref{xrand}) in the arguments of each of the trigonometric expansion terms in (\ref{deltaj}) 
produce uniformly distributed random variables $x_{i}$. This yields a discretized It\^o type 
equation
\begin{eqnarray}
\delta^{(j-1)}=f^{(j-1)}_{\delta}-\sum_{p}C_{p}^{(j-1)}
\sin\left(\vec m^{(j-1)}_{p}\vec x+\varphi^{(j-1)}_{p}\right),
\label{deltajx}
\end{eqnarray}
where $\delta_{1}^{(j)}$ and $\delta_{2}^{(j)}$ are also functions of $x$, produced 
by the corresponding expansions for $\delta_{n}^{(j)}$ and $\delta_{n-1}^{(j)}$. For 
a given $j$, these functions introduce the fluctuations from the lower levels of the 
hierarchy into (\ref{deltajx}).

It is possible to use the systems of expansions (\ref{deltaj}) directly to produce 
the probability distributions for each of the $\delta^{(j)}$s. However, it is more 
illustrative to use instead a simple physical approximation, in which the $j$th level 
fluctuations $\delta_{n}^{(j)}$ and $\delta_{n-1}^{(j)}$ are treated as independent 
random variables $\delta_{1}^{(j)}$ and $\delta_{2}^{(j)}$, distributed according to 
$P_{\delta}^{(j)}$. This allows to follow the accumulation of the fluctuations of the 
final distributions for $\delta^{(0)}$ starting from the fluctuations at the regular 
level, $\delta^{(r)}$.
%%%%%%%%%%%%%%%%%%%%%%%%%%%%%%%%%%%%%%%%%%%%%%%%%%%%%%%%%%%%%%%%%%%%%%%%%%%%%%%
\begin{figure}[tbp]
\begin{center}
\includegraphics{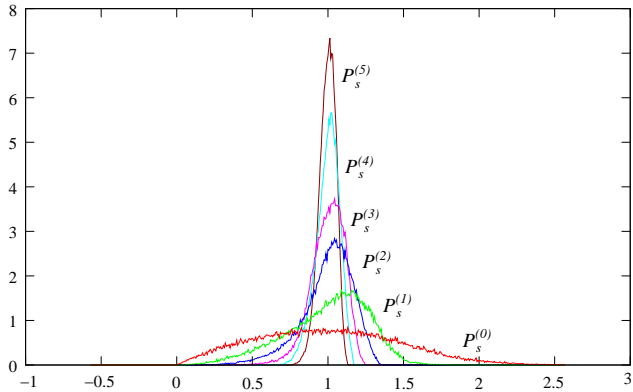}
\end{center}
\caption{The development of the full scale spectral fluctuations for the nearest
neighbor separations ($s^{(0)}$, shown in red) for the fully connected quadrangle 
graph studied also in \cite{QGT} for $r=6$. One can follow the appearance of the 
Wignerian type distribution at the physical level from the distributions $P^{(j)}(s)$.}
\label{Fig.4}
\end{figure}
%%%%%%%%%%%%%%%%%%%%%%%%%%%%%%%%%%%%%%%%%%%%%%%%%%%%%%%%%%%%%%%%%%%%%%%%%%%%%%%%%
At the regular level, $j=r$, the fluctuations $\delta^{(r)}$ are described by an 
expansion structurally identical to (\ref{deltax}) so the corresponding distribution 
$P_{\delta}^{(r)}$ is self-contained and has the same functional form as (\ref{distreg}). 
Knowing $P_{\delta}^{(r)}$ allows to obtain $P_{\delta}^{(r-1)}$, and so on, up to the 
distribution $P_{\delta}^{(0)}$, which applies to the fluctuations of the physical 
momentum eigenvalues. 
With the independent $\delta_{1}^{(j)}$ and $\delta_{2}^{(j)}$ the transition from 
$P_{\delta}^{(j)}$ to $P_{\delta}^{(j-1)}$ is direct. Proceeding as in (\ref{deltax})
-(\ref{Fk}) yields the result 
\begin{equation}
P_{\delta}^{(j-1)}(\delta^{(j-1)})=\int dke^{ik\delta^{(j-1)}}
\left\langle F^{(j-1)}_{\delta}(k)\right\rangle_{\Omega^{(j-1)}},
\label{pjminusone}
\end{equation}
where $F^{(j-1)}_{\delta}(k,\delta^{(j)})$ is obtained as (\ref{Fk}) for the series 
(\ref{deltajx}), and $\langle *\rangle_{\Omega^{(j)}}$ represents averaging over the 
``disorder'' produced by the separators $\delta_{1}^{(j)}$ and $\delta_{2}^{(j)}$ of 
$j$th level using the weight function
\begin{eqnarray}
\Omega^{(j-1)}=e^{ikf^{(j-1)}_{\delta}\left(\delta_{1}^{(j)},\delta_{2}^{(j)}\right)}
P_{\delta}^{(j)}\left(\delta^{(j)}_{1}\right)P_{\delta}^{(j)}\left(\delta^{(j)}_{2}\right)
\label{ga}
\end{eqnarray}
Similar considerations produce the distributions for the nearest neighbor separations, 
the form factor, etc.

\section{Discussion}

The explicit periodic orbit representations of the individual momentum eigenvalues 
$k_{n}$ of quantum graphs \cite{Fabula} provide a regular analytical method 
of describing the statistical properties of their spectra within the standard periodic 
orbit theory framework.

As shown in \cite{Fabula}, solving the spectral problem for a generic quantum 
graph goes beyond the conventional Gutzwiller's theory approach. The additional information 
supplied by the auxiliary separating sequences $\hat k_{n}^{(j)}$ allows to produce the exact 
physical spectrum $k_{n}=\hat k_{n}^{(0)}$, to reveal the hierarchical structure of 
the spectral fluctuations and then to unfold the exact probability distributions for spectral 
statistics associated with each sequence $\hat k_{n}^{(j)}$.

Among other things, the expansions (\ref{deltax}) also provide a physical understanding 
about the origins of the universal behavior of spectral characteristics based on the 
periodic orbit theory. It is known that the terms of lacunary trigonometric series behave 
as weakly dependent random variables (see e.g. \cite{Proxorov,Revesz} and the references 
therein). This fact allows to establish a number of universal features for the asymptotic 
distributions of their sums, such as convergence to Gaussian distribution with a specific 
variance (also conjectured to be the universal probability distribution profile for the
$\delta^{(0)}_{n}$ distribution of generic quantum chaotic systems \cite{ABS}). 
In addition, the build up of the distributions $P^{(j)}$ along the levels of the hierarchy 
can lead to the appearance of other universal, (albeit more complex, e.g. Wignerian, see 
Fig.~4 and \cite{BGS}) profiles.

The advantage of this result is twofold. On the one hand, since the spectral properties of 
sufficiently complex quantum graphs are well described by the RMT \cite{QGT}, the proposed 
approach provides a mechanism to follow the build up of the universal distributions as they 
appear at the top level of the spectral hierarchy. On the other hand, the analysis of a given
system can be made as detailed as necessary (by using higher order approximations in each of
the expansions (\ref{deltaj})), (\ref{deltajx}), so the individual features are not overlooked
behind broad universality.

%%%%%%%%%%%%%%%%%%%%%%%%%%%%%%%%%%%%%%%%%%%%%%%%%%%%%%%%%%%%%%%%
Work supported in part by the Sloan and Swartz Foundations. 
%%%%%%%%%%%%%%%%%%%%%%%%%%%%%%%%%%%%%%%%%%%%%%%%%%%%%%%%%%%%%%%%

\end{document}